\documentclass[letterpaper]{article} 
\usepackage[submission,draft]{aaai2027}
\usepackage[hyphens]{url}  
\usepackage{graphicx} 
\usepackage{natbib}  
\usepackage{caption} 
\usepackage{algorithm}
\usepackage{algorithmic}
\usepackage{datetime2}
\RequirePackage{fancyhdr}
\makeatletter
\def\showauthors@on{T}
\makeatother
\usepackage{newfloat}
\usepackage{listings}
\DeclareCaptionStyle{ruled}{labelfont=normalfont,labelsep=colon,strut=off} 
\floatstyle{ruled}
\newfloat{listing}{tb}{lst}{}
\floatname{listing}{Listing}

\usepackage{booktabs}

\usepackage{amsmath}
\usepackage{amsfonts}
\usepackage{siunitx}
\usepackage{makecell}
\usepackage{multirow}
\usepackage{graphicx}
\usepackage[table]{xcolor}
\usepackage{enumitem}

\definecolor{tableblue}{RGB}{237,245,249}
\usepackage{datetime2}
\RequirePackage{fancyhdr}

\title{Think Thrice Before Reranking: Multi-perspective Evidence and Reasoning Integration for Text Reranking}
\author{
Lijun Liu$^{1}$,
Zhengzong Chen$^{1}$\thanks{Corresponding authors.},
Wenyan Li$^{1}$,
Yuanyuan Zhao$^{1}$,
Fei Huang$^{1}$\\
$^{1}$\textnormal{Honor Device Co., Ltd}\\
}
\affiliations{
    Correspondence: chenzhengzong@honor.com
}

\makeatletter

\def\twodigits#1{\ifnum#1<10 0\fi\the#1}

\fancypagestyle{firstpage}{
  \fancyhf{}
  \fancyfoot[C]{\thepage}

  \lhead{\includegraphics[height=20pt]{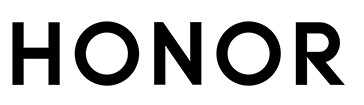}}
  \rhead{\textcolor[HTML]{4D4D4D}{\DTMtoday}}
  \renewcommand{\headrulewidth}{1pt}
  \renewcommand{\headrule}{%
    {\color[HTML]{4D4D4D}%
    \hrule\@height\headrulewidth\@width\headwidth\vskip-\headrulewidth}%
  }
  \setlength{\headheight}{1pt}   
  \setlength{\headsep}{-5pt}       
}
\makeatother
\renewcommand{\headrulewidth}{1pt}
\begin{document}
\thispagestyle{firstpage}
\maketitle
\pagestyle{fancy}
\begin{abstract}
Reasoning-based reranking with Large Language Models (LLMs) has shown promising improvements in text ranking. However, current methods predominantly rely on a single reasoning trajectory, resulting in rankings that are susceptible to reasoning errors and inherently constrained in modeling the multifaceted signals underlying document relevance. To resolve this dilemma, we propose \textbf{MERIT-Rank} (Multi-perspective Evidence and Reasoning Integration for Text Reranking), a framework that models complementary reasoning trajectories to improve reranking robustness. MERIT-Rank formulates a Multi-Trajectory Reasoning Space (MTRS) that evaluates query–document relevance from multiple perspectives and introduces a joint reranker that consolidates these reasoning paths into a unified ranking decision. We further develop Progressive Rank Policy Optimization (PRPO), a progressive training framework that stabilizes reasoning trajectories while continually improving ranking quality through staged optimization objectives. Experiments on both reasoning‑intensive and traditional retrieval benchmarks show that MERIT‑Rank consistently achieves superior performance over competitive baselines. The 4B model notably outperforms most 7B and even 32B rerankers on BRIGHT.

\end{abstract}


\section{Introduction}

Information Retrieval (IR) systems typically adopt a two-stage architecture consisting of an initial retrieval step followed by a reranking stage that refines the candidate list according to query–document relevance. 
In response to more complex retrieval scenarios~\cite{BRIGHT,followir}, reranking methods have been developed under pointwise~\cite{rankt5}, pairwise~\cite{qin2024larg_pairwise}, setwise~\cite{zhuang2024setwise}, and listwise formulations~\cite{weller2025rank1,zhang2025rearank}.
Among them, reasoning-based listwise reranking~\cite{rank-k,liu2025reasonrank} has shown strong performance. Incorporating test-time reasoning~\cite{guo2025deepseekr1} allows it to perform fine-grained relevance assessment and capture deeper latent semantic query–document dependencies.

Despite their effectiveness, existing reasoning-based reranking methods~\cite{liu2025reasonrank, zhang2025rearank} rely heavily on a single reasoning trajectory to derive the final ranking. The model first generates reasoning and subsequently produces the ranking conditioned on that reasoning. 
Consequently, the ranking becomes vulnerable to reasoning quality: errors in the reasoning chain can propagate through autoregressive generation and degrade the final ranking.
Furthermore, query–document relevance in real-world retrieval scenarios is inherently multi-faceted, involving diverse signals such as semantic alignment, user intent satisfaction, and evidence support within documents. 
A single reasoning trajectory captures only a limited subset of available signals, thereby constraining the robustness and generalization of reasoning-based rerankers across diverse retrieval tasks.


\begin{figure}[t]
\centering
\includegraphics[width=\columnwidth]{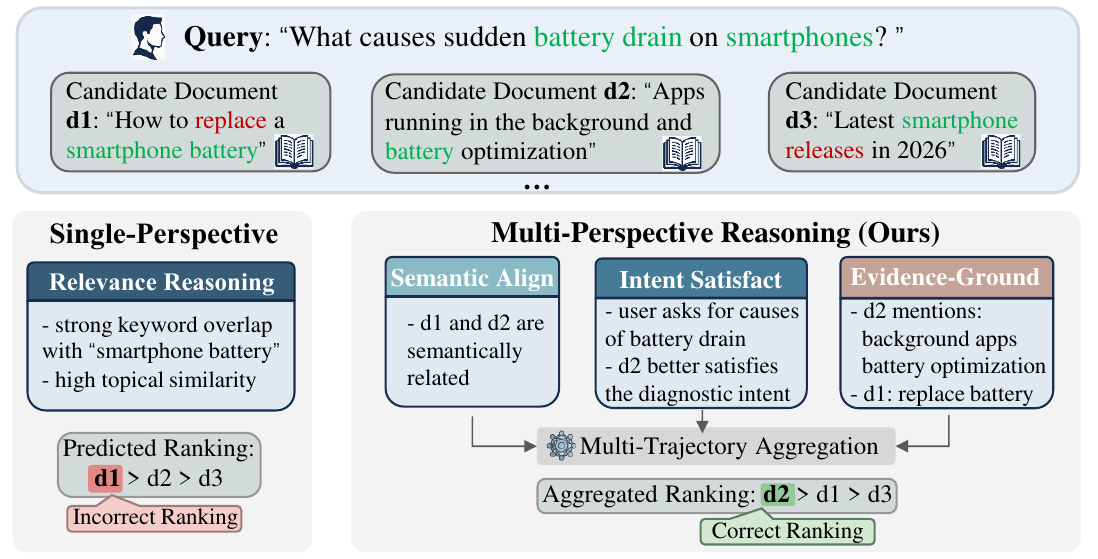}
\caption{Illustration of different reasoning paradigms in reranking. While single-perspective reasoning may struggle to capture implicit user intent, our approach aims to mitigate this by incorporating diverse and complementary evidence.
}
\label{fig:efficiency}
\vspace{-10pt}
\end{figure}

\begin{figure*}[tb]
  \centering
  \includegraphics[width=0.95\linewidth]{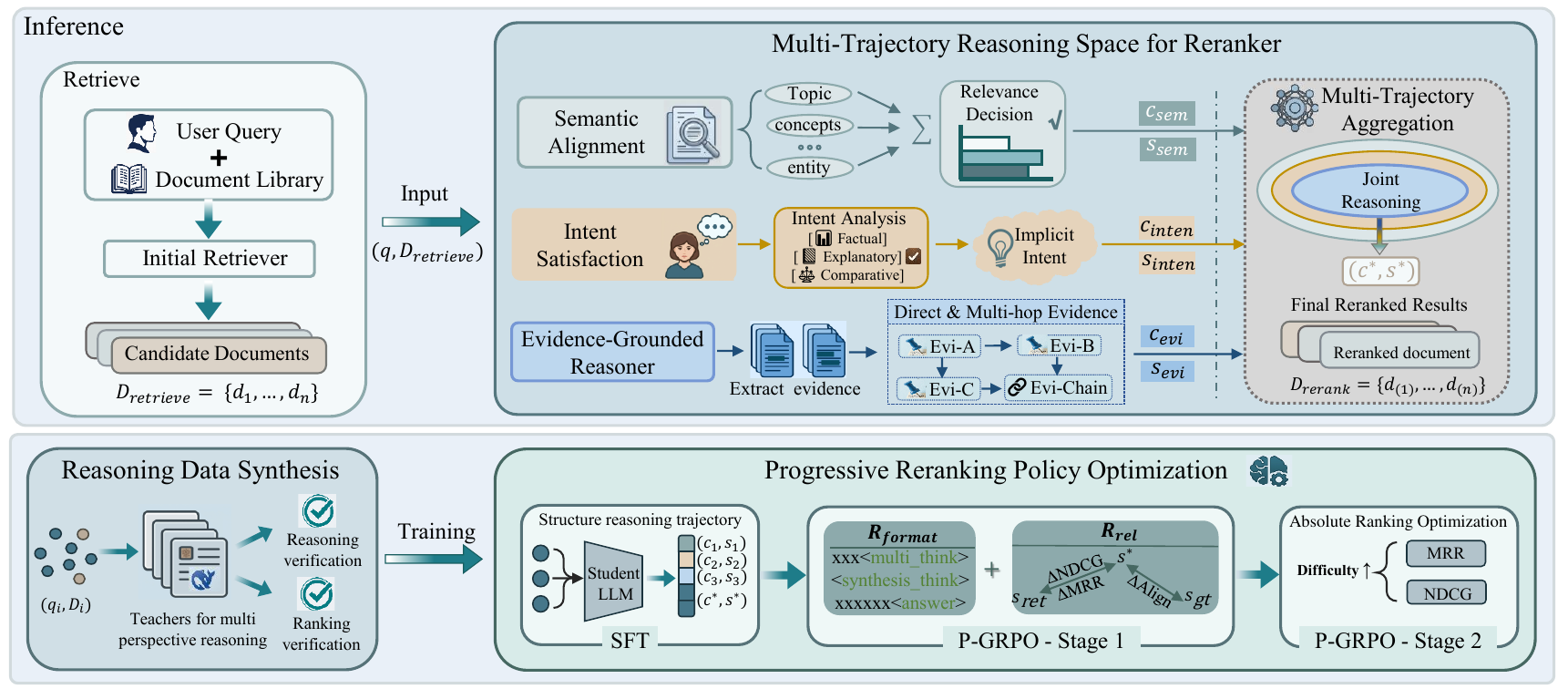}
  \caption{
  Overview of MERIT-Rank. Given a query and initially retrieved candidates, the framework constructs a Multi‑Trajectory Reasoning Space to generate complementary reasoning trajectories, which are then aggregated via synthesized reasoning for the final ranking. To effectively learn in this space, MERIT-Rank employs reasoning-augmented data synthesis and a progressive optimization strategy that combines SFT with staged policy optimization.
  }
  \label{fig:overview}
  \vspace{-2mm}
\end{figure*}

To overcome these limitations, we introduce MERIT-Rank (Multi-perspective Evidence and Reasoning Integration for Text Reranking), a novel and robust framework that improves reranking performance by explicitly modeling complementary reasoning trajectories from multiple perspectives and aggregating them into a unified ranking decision. The key intuition behind MERIT-Rank is that diverse reasoning perspectives can provide highly heterogeneous yet complementary relevance signals. To this end, we construct a multi-trajectory reasoning space that explores various perspectives, including semantic alignment, intent satisfaction, and evidence-grounded reasoning, as shown in Figure~\ref{fig:overview}. These complementary trajectories capture richer relevance signals and mitigate bias from any single reasoning perspective.

Building upon this expanded reasoning space, we further propose a multi-trajectory joint reranker that seamlessly unifies several reasoning and ranking within a single generative framework.
To avoid the proliferation of models typical of conventional ensemble methods~\cite{bruch2023analysis}, we consolidate multi‑perspective reasoning and ranking aggregation within a single model. The model generates a reasoning chain and an intermediate ranking for each perspective, and subsequently aggregates them to derive the final prediction. Guided by this joint reasoning paradigm, the reranker integrates cross‑perspective evidence in a structured manner, enhancing interpretability while producing more stable and reliable final ranking decisions.

Despite its advantages, training a multi‑ trajectory reasoning model remains challenging, largely due to the extended reasoning sequences and the difficulty of learning stable ranking policies.
To overcome these challenges, we propose Progressive Rank Policy Optimization (PRPO), a curriculum‑style training strategy that unifies Supervised Fine‑Tuning (SFT) with Reinforcement Learning (RL) objectives of progressively increasing difficulty.
The model is first trained via SFT on structured multi‑trajectory reasoning data to establish a consistent reasoning format. RL objectives are then introduced in stages: initially encouraging improvements over the baseline retrieval ranking, and subsequently optimizing absolute ranking metrics such as MRR and NDCG. The PRPO paradigm supports continuous improvement, mitigating the risk of convergence to suboptimal local minima while further improving ranking effectiveness.

Extensive experiments on both reasoning-intensive and semantically relevant IR benchmarks demonstrate that MERIT-Rank consistently achieves state-of-the-art reranking performance while exhibiting strong out‑of‑domain generalization.
Notably, MERIT‑Rank achieves competitive performance even with a relatively small model size, reflecting substantial parameter efficiency and further underscoring the effectiveness of our proposed structured multi‑perspective reasoning framework.
Moreover, while achieving the same performance as SOTA methods, MERIT‑Rank efficiently requires fewer sliding windows, generating fewer tokens and thereby achieving lower overall inference latency.

Our contributions can be summarized as follows:
\begin{itemize}
\item We propose MERIT-Rank, a novel multi-perspective reasoning framework for reranking that mitigates the limitations of single-trajectory reasoning by constructing a complementary multi-trajectory reasoning space.
\item We develop a multi-trajectory joint reranker that jointly performs perspective-specific and comprehensive reasoning within a unified framework to derive the final ranking.
\item We propose PRPO, a curriculum‑driven SFT– RL training paradigm that gradually shifts from format learning to relative ranking and ultimately to absolute ranking optimization, yielding consistent gains in ranking quality.

\item Extensive experiments consistently show that MERIT-Rank achieves promising performance, strong out-of-domain generalization, and competitive efficiency across reasoning-intensive and semantically relevant IR tasks.
\end{itemize}

\section{Related Work} 

\paragraph{LLM-based Rerankers}

LLMs have significantly improved text reranking, outperforming earlier encoder-based models such as BERT~\cite{liu2024information}. Existing methods generally follow {pointwise, pairwise, and listwise paradigms}.
{Pointwise} methods independently score each query--document pair, as in monoBERT~\cite{monobert} and generative rerankers such as MonoT5~\cite{monot5}, with later LLM-based extensions~\cite{liang2022holistic,sachan2022improving,rankllama,liu2024demorank}. These methods are efficient but ignore interactions among candidate documents.
{Pairwise approaches} rank documents through pair comparisons, such as duoBERT~\cite{nogueira2019duobert} and DuoT5~\cite{pradeep2021duot5}, with later improvements in comparison strategies~\cite{qin2024larg_pairwise,PRP-Graph}. While providing stronger ranking signals, they incur higher computational cost.
{Listwise rerankers} jointly evaluate multiple documents to infer their ranking order. Prompt-based methods such as RankGPT~\cite{rankgpt} and related approaches~\cite{ma2023zero,zhang2024two} demonstrate strong performance, and subsequent work explores more efficient listwise or setwise ranking strategies~\cite{rankzephyr,listt5,tourrank,fan2025llm,zhuang2024setwise}. 


\paragraph{Reasoning-based Reranking}

Recent work incorporates \textbf{explicit reasoning} into reranking models for complex retrieval tasks. Prompt-based approaches such as JudgeRank~\cite{judgerank} and InsertRank~\cite{insertrank} use LLM reasoning during ranking, while other studies introduce reasoning through training. For instance, Rank1~\cite{weller2025rank1} and Rank-K~\cite{rank-k} distill reasoning chains from stronger models, and reinforcement learning approaches such as Rank-R1~\cite{rank-r1}, REARANK~\cite{zhang2025rearank}, and ReasonRank~\cite{liu2025reasonrank} optimize reasoning-aware ranking policies. These methods show strong results on reasoning-intensive benchmarks such as BRIGHT~\cite{BRIGHT}, but typically rely on a \textbf{single reasoning trajectory}, which can lead to hallucination and error accumulation.

\section{Method}

We propose {MERIT-Rank} (Multi-perspective Evidence and Reasoning Integration for Text Reranking)
to mitigate the over-reliance on single reasoning trajectories, as shown in Figure~\ref{fig:overview}. Specifically, we first construct a complementary Multi-Trajectory Reasoning Space (MTRS), generating diverse reasoning chains from multiple reasoning perspectives. Subsequently, we introduce a progressive joint optimization strategy to train a multi-trajectory reranker, integrating signals across various paths for comprehensive reasoning.

\subsection{Task Formulation}
Given a query $q$ and retrieved candidate document $D = \{d_1, d_2, ..., d_n\}$, typical reasoning-based reranking approaches leverage LLMs to generate reasoning and the reranked sequence autoregressively. Specifically, model learns the following conditional probability distribution:
\begin{equation}
P_\theta(y|q,D) = \prod_{t=1}^{T} P_\theta(y_t \mid q,D,y_{<t}),
\end{equation}
where the output sequence $y = (c, s)$ comprises two segments: a {reasoning chain} $c = (y_1, \ldots, y_k)$ elucidating query-document relevance, followed by the final {ranking sequence} $s = (y_{k+1}, \ldots, y_T)$.

Since the ranking sequence $s$ is conditioned on the previously generated reasoning chain $c$, the entire generation process can be decomposed as:
\begin{equation}
P_\theta(y|q,D) = P_\theta(c|q,D)\, P_\theta(s|q,D,c).
\end{equation}
This equation indicates that the model first generates the reasoning trajectory $c$ based on $q$ and $D$, and subsequently produces the ranking sequence $s$ conditioned on it. As a result, existing reasoning-based rerankers rely heavily on the single reasoning trajectory. Once a local error occurs during reasoning, it propagates through the autoregressive decoding process, ultimately misleading the final ranking.

\subsection{Complementary MTRS}
\label{method:comple}


In real-world ranking scenarios, query--document relevance may depend on multiple aspects, such as semantic matching, intent satisfaction, and evidence support. However, a single reasoning trajectory often focuses on only part of these signals, which may lead to incomplete relevance estimation in complex ranking tasks. To address this issue, we construct a complementary multi-trajectory reasoning space to explore diverse reasoning paths and enrich relevance modeling.

\subsubsection{Complementary Reasoning Motivation}


Different reasoning perspectives emphasize distinct aspects of query--document relevance. For example, one trajectory may focus more on semantic similarity, while another emphasizes factual evidence or intent alignment. As a result, relying on a single trajectory may produce biased or incomplete ranking decisions. Motivated by this observation, we investigate whether aggregating multiple reasoning trajectories can provide more comprehensive relevance signals for reranking. 

Formally, we denote the ranking sequence generated from the $i$-th reasoning perspective as $s_i$, where
$s_i \sim f_i(s \mid q, D)$. 
The final ranking result is obtained by aggregating the outputs from multiple reasoning perspectives:
\begin{equation}
F(q, D) = P_\theta(f_1(q,D), f_2(q,D), ..., f_n(q,D)).
\end{equation}
The expected error of each function is defined as:
\begin{equation}
\epsilon_i = \mathbb{E}[L(f_i(q,D))],
\end{equation}
where $L(\cdot)$ denotes the ranking loss function.

Although these reasoning trajectories are generated from the same underlying LLM and therefore do not satisfy the independence assumption commonly adopted in traditional ensemble learning~\cite{krogh1994neural}, prior studies~\cite{hu-etal-2025-dipper} on homogeneous LLM ensembles have shown that different prompts can still induce diverse intermediate reasoning processes and decision patterns from the same model. As a result, different reasoning trajectories may focus on different aspects of query--document relevance and provide complementary relevance evidence. Therefore, aggregating multiple reasoning trajectories, rather than relying on a single reasoning path, has the potential to improve the robustness and generalization ability of reranking models.

\subsubsection{Reasoning Perspective Modeling}

Based on the above observations, we construct a Multi-Trajectory Reasoning Space, which introduces heterogeneous perspectives to generate multiple complementary reasoning trajectories. Specifically, we design distinct reasoning perspectives that assess query–document relevance from different dimensions.

\textbf{Semantic Alignment Reasoning}. 
This perspective focuses on the semantic correspondence between the query and the retrieved candidate documents. It encourages the model to evaluate topic coherence, key concept correspondence, and entity-level consistency in order to determine relevance.

\textbf{Intent Satisfaction Reasoning}. 
In real-world retrieval scenarios, a document may be topically related to the query yet still fail to satisfy the user’s underlying information need. This perspective first infers the implicit intent of the query (e.g., factual, explanatory, or comparative) and then evaluates whether the document adequately fulfills this need.

\textbf{Evidence-Grounded Reasoning}. 
This perspective grounds the model’s reasoning in explicit textual evidence from the document, thereby reducing hallucinated reasoning. 
It specifically emphasizes identifying supporting facts or key passages as evidence and constructing logical chains that connect this evidence to the original query. 
The reasoning may involve direct evidence matching or multi-hop evidence aggregation, depending on query complexity.

\subsection{Progressive Reranker Optimization}

To enable ranking within the multi-trajectory reasoning space, we first construct training data that contains multiple complementary reasoning trajectories and a synthesized reasoning chain, providing richer supervision than traditional relevance labels. 
Building on this foundation, we propose a unified progressive training framework that integrates Supervised Fine-Tuning (SFT) with increasingly difficult Reinforcement Learning (RL) objectives, thereby progressively improving both reasoning quality and ranking performance.



\subsubsection{Multi-Trajectory Data Synthesis}
\label{sec:dataset_sys}

In traditional information retrieval datasets, each training instance is typically represented as a triplet $(q, D, D_y)$, where $q$ denotes a query, $D = \{d_1, d_2, \ldots, d_n\}$ represents the set of candidate documents associated with the query, and $D_y \subseteq D$ denotes the subset of documents that are labeled as relevant to $q$. However, in the context of multi-trajectory reasoning, the training data must incorporate diverse reasoning paths alongside synthesized supervision signals to guide the model in making perspective-aware ranking decisions.

To this end, we employ DeepSeek-R1 \cite{guo2025deepseekr1} as teacher to generate both sets of supervision signals. First, it generates multiple reasoning trajectories and initial ranked sequences in parallel, denoted as $\mathcal{R}_{gold}$. Based on $\mathcal{R}_{gold}$, DeepSeek-R1 is further utilized to synthesize a consolidated reasoning chain and derive a gold ranking, collectively denoted as $\mathcal{R}^*_{gold}$. To ensure data usability, if any of these generated outputs lack the requisite `<think>` tags or exhibit formatting errors, GPT~\cite{achiam2023gpt} is uniformly employed to regenerate and correct them. Consequently, final training instances are constructed as the following tuple:
\begin{equation}
(q, D, D_y, \mathcal{R}_{gold}, \mathcal{R}^*_{gold}).
\end{equation}
To ensure the quality of the synthesized reasoning and ranking, we adopt a dual-path verification mechanism to filter out samples that fail to satisfy either validation criterion:
1) {Reasoning validity}: The synthesized ranking should outperform the rankings produced by any individual reasoning perspective, as measured by NDCG@10.
2) {Ranking validity}: The synthesized ranking sequence within $\mathcal{R}^*_{gold}$ must place the relevant documents $D_y$ at the top positions.

\begin{table*}[t]
\centering
\small
\caption{The results (NDCG@10) on BRIGHT benchmark. All models rerank ReasonIR-retrieved top-100 passages. }
\label{tab: bright}
\resizebox{\textwidth}{!}{
\begin{tabular}{lcccccccccccccc}
\toprule
\multirow{2}{*}{\textbf{Methods}} & \multicolumn{7}{c}{\textbf{StackExchange}} & \multicolumn{2}{c}{\textbf{Code}} & \multicolumn{3}{c}{\textbf{Theorem-based}} & \multirow{2}{*}{\textbf{Avg}} \\
\cmidrule(lr){2-8} \cmidrule(lr){9-10} \cmidrule(lr){11-13}
 & Bio & Earth & Econ & Psy & Rob & Stack & Sus & Pony & Leet & AoPS & TheoQ & TheoT &  \\

\midrule
ReasonIR-8B
& 43.5 & 43.0 & 32.7 & 39.6 & 20.8 & 31.0 & 27.3 & 19.6 & {31.7} & 7.4 & 33.9 & 36.7 & 30.6 \\
JudgeRank-8B
& 37.1 & 27.2 & 19.2 & 28.6 & 11.6 & 19.9 & 22.5 & 10.2 & 10.2 & 3.6 & 22.9 & 29.4 & 20.2 \\

Rank-R1-14B
& 44.5 & 38.7 & 27.4 & 37.1 & 23.1 & 27.8 & 36.8 & 19.2 & 21.3 & 8.8 & 31.7 & 39.5 & 29.7 \\

Rank1-32B
& 42.4 & 38.0 & 25.4 & 34.9 & 17.1 & 23.8 & 31.2 & \textbf{41.0} & 12.2 & 4.8 & 29.3 & 40.0 & 28.3 \\

REARANK-7B
& 46.9 & 40.3 & 30.7 & 40.8 & 27.2 & 26.1 & 36.3 & 22.8 & 30.6 & 7.3 & 32.3 & 39.9 & 31.8 \\

Rank-K-32B
& 50.6 & 39.8 & 30.1 & 43.5 & 26.6 & 29.9 & 35.2 & 22.8 & 27.2 & 7.6 & 37.1 & 41.0 & 32.6 \\

ReasonRank-7B
& 56.7 & 47.8 & {35.1} & 47.8 & {31.2} & 32.5 & 40.9 & 25.0 & 23.2 & 7.7 & \underline{39.5} & 41.8 & 35.7 \\

ERANK-4B
& 42.1 & 42.5 & 26.3 & 36.4 & 20.8 & 27.3 & 33.2 & 21.8 & {31.7} & {10.9} & 32.8 & 40.6 & 30.5 \\

ERANK-14B
& 46.6 & 42.5 & 25.2 & 37.3 & 19.6 & 30.2 & 34.6 & {25.6} & {31.9} & 10.5 & 32.4 & \underline{45.0} & 31.8 \\

ERANK-32B & 49.3 & 43.4 & 28.4 & 36.8 & 20.8 & 32.8 & 34.6 & 22.3 & \underline{36.0} & \underline{11.3} & 34.4 & {43.5} & 32.8 \\

\rowcolor{tableblue}
MERIT-Rank-4B & \underline{59.1} & \underline{50.6} & 35.0 & \underline{50.7} & \underline{32.4} & {34.5} & \underline{44.8} & 24.6 & 28.4 & 8.7 & 37.8 & 33.0 & {36.6} \\

\rowcolor{tableblue}
MERIT-Rank-7B & {58.8} & {49.5} & \underline{37.5} & {49.5} & 30.6 & \underline{35.1} & {42.3} & 22.3 & 31.0 & 8.4 & {39.0} & 41.7 & \underline{37.1} \\

\rowcolor{tableblue}
MERIT-Rank-32B & \textbf{61.1} & \textbf{50.9} & \textbf{39.2} & \textbf{52.2} & \textbf{33.4} & \textbf{39.6} & \textbf{46.8} & \underline{25.8} & \textbf{36.5} & \textbf{12.2} & \textbf{40.4} & \textbf{45.6} & \textbf{40.3} \\

\bottomrule
\end{tabular}
}
\end{table*}

\subsubsection{Multi-Trajectory Joint Reranker}

To better exploit the supervision signals in the training data, we introduce a multi-trajectory joint reranker. The key idea is to explicitly model multiple reasoning perspectives within a unified context, thereby enabling collaboration between {multi-perspective reasoning} and {synthesized reasoning}.

First, given the multi-perspective set $\mathcal{P}=\{p_{sem}, p_{intent}, p_{evi}\}$, the proposed reranker generates a reasoning trajectory $c_p$ and a corresponding ranking sequence $s_p$ for each perspective $p\in\mathcal{P}$ in turn. And the entire multi-perspective reasoning is formulated as follows:
\begin{equation}
P_\theta(\mathcal{R} \mid q, D, \mathcal{P}),
\end{equation}
where $ \mathcal{R} = \{R_p\}_{p \in \mathcal{P}} $ denotes reasoning results across all perspectives, and each result $R_p = (c_p, s_p)$ consists of a reasoning chain and its corresponding ranking sequence.

Next, the reranker integrates evidence from all perspectives to produce the final joint result:
\begin{equation}
P_\theta(\mathcal{R}^* \mid q, D, \mathcal{P}, \mathcal{R}),
\end{equation}
where $\mathcal{R}^* = (c^*, s^*)$, with $c^*$ denoting the aggregated reasoning chain that integrates evidence from multiple perspectives, and $s^*$ representing the final ranking sequence predicted based on the comprehensive information within $c^*$.

In our specific implementation, the above reasoning and ranking results are organized into a unified sequence
$y = (y_1, y_2, \ldots, y_T) = (\mathcal{R}, \mathcal{R}^*)$, and the model generates each token in an autoregressive manner:
\begin{equation}
P_\theta(y \mid q, D, \mathcal{P}) = \prod_{t=1}^{T} P_\theta(y_t \mid q, D, \mathcal{P}, y_{<t}).
\end{equation}
In this way, the reranker jointly models multi-perspective reasoning and the final ranking decision within a unified framework. During inference, it sequentially performs multi-perspective reasoning, perspective-level ranking, multi-trajectory aggregation, and final ranking prediction, yielding more consistent and reliable ranking results.

\subsubsection{Progressive Rank Policy Optimization}

However, directly training on high-difficulty multi-trajectory reasoning tasks is inefficient. Long reasoning processes often lead to reasoning collapse, where the model bypasses multi-perspective reasoning and directly outputs rankings. Moreover, multi-trajectory reasoning produces long chains, making direct policy exploration inefficient.
To address these challenges, we propose Progressive Rank Policy Optimization (PRPO), which starts with easier tasks and gradually introduces more complex objectives to improve training stability.


Based on the supervised data constructed in the previous section
, we first conduct Supervised Fine-Tuning (SFT) to 
learn the basic structured reasoning process:
\begin{equation}
\mathcal{L}_{SFT} = - \mathbb{E}{(\mathbf{x}, y_{gold})}
\sum_{t=1}^{T} \log P_\theta (y_t \mid \mathbf{x}, y_{<t}),
\end{equation}
where $\mathbf{x}=(q,D,\mathcal{P})$ denotes the complete input context information. The supervised sequence $y_{gold} = (\mathcal{R}_{gold}, \mathcal{R}^*_{gold})$ contains both multiple complementary reasoning trajectories and the final aggregated reasoning trajectory. 

Considering that SFT learns by imitating reasoning trajectories and thus is sensitive to data distribution, we further introduce a reinforcement learning stage. Building on GRPO~\cite{shao2024deepseekmath}, we propose a two-stage Progressive GRPO (P-GRPO) to optimize the aggregated ranking sequence in the multi-trajectory reasoning space.

The first-stage GRPO establishes a basic reasoning policy, focusing on format and preliminary ranking ability. The format reward $R_{format}$ enforces a multi-level reasoning structure to prevent shortcut behavior: a binary reward of $1$ is assigned if the output contains multi-trajectory reasoning beginning with \texttt{<multi\_think>}, synthesized reasoning with \texttt{<synthesis\_think>}, and the final ranking with \texttt{<answer>}; otherwise, it is penalized with a value of $-1$.
The ranking ability is optimized via a relative reward, encouraging the synthesized ranking $s^*$ to outperform the initial retrieval sequence $s_{retrieve}$ in NDCG and MRR while approaching the ground-truth ranking $s_{gt}$:
\begin{equation}
R_{rel}=R_{\Delta NDCG} + R_{\Delta MRR} + R_{\Delta Align},
\end{equation}
where $R_{\Delta NDCG}$ denotes the NDCG@10 improvement, and $Align(\cdot)$ measures the Rank-Biased Overlap (RBO)~\cite{rbo} between the synthesized ranking $s^*$ and $s_{gt}$. Thus, the total combined reward for the first-stage GRPO is formally defined as:
\begin{equation}
R_1 = R_{format} + R_{rel}.
\end{equation}

The second-stage GRPO further increases difficulty by introducing a suite of rigorous absolute ranking metrics, including MRR and NDCG, to better encourage the generation of stable and optimal re-ranking sequences:
\begin{equation}
R_2 = R_1 + \lambda_1 \, MRR + \lambda_2 \, NDCG, 
\end{equation}
where $\lambda_i$ is the weight of each component (default $1$). This progressive scheme—shifting from relative improvement to absolute optimization—first learns stable gains over the initial retrieval ranking and then promotes globally optimal ranking, improving GRPO stability and convergence.

\section{Experiment}


\begin{table*}[t]
\centering
\small
\setlength{\tabcolsep}{5pt}
\caption{The results (NDCG@10) on TREC-DL and BEIR benchmarks. All models rerank BM25 top-100 passages.}
\label{tab: BEIR}
\resizebox{\textwidth}{!}{
\begin{tabular}{lccccccccccc}
\toprule
\multirow{2}{*}{\textbf{Methods}} & \multicolumn{3}{c}{\textbf{TREC}} & \multicolumn{8}{c}{\textbf{BEIR}} \\
\cmidrule(lr){2-4} \cmidrule(lr){5-12}
 & Avg & DL19 & DL20 & Avg & Covid & NFCorpus & DBPedia & SciFact & Signal & News & Robust04 \\
\midrule

\multicolumn{12}{l}{\textit{Non-reasoning reranker}} \\

BM25
& 49.3 & 50.6 & 48.0 & 43.3 & 59.5 & 30.8 & 31.8 & 67.9 & 33.1 & 39.5 & 40.7 \\
Qwen$_{2.5}$-7B
& 65.5 & 68.3 & 62.7 & 50.1 & 77.7 & 37.4 & 39.8 & 70.8 & 31.7 & 43.2 & 50.0 \\
GPT$_{3.5}$
& 64.4 & 65.8 & 62.9 & 51.3 & 76.7 & 35.6 & 44.5 & 70.4 & 32.1 & 48.9 & 50.6 \\
RankMistral
& 69.9 & 71.7 & 68.1 & 46.0 & 78.0 & 33.1 & 37.7 & 66.2 & 30.0 & 37.1 & 39.5 \\
RankZephyr
& 72.3 & 73.9 & 70.6 & 54.2 & 83.5 & 38.4 & 44.3 & 75.2 & 31.4 & 52.4 & 54.2 \\
GPT$_4$
& {73.1} & \underline{75.6} & 70.6 & {55.8} & {85.5} & 38.5 & \underline{47.1} & 75.0 & 34.4 & {52.9} & 57.6 \\
\midrule
\multicolumn{12}{l}{\textit{Reasoning reranker}} \\
Qwen$_3$-32B
& 71.6 & 73.1 & 70.0 & 54.1 & 83.9 & 36.3 & 45.4 & 71.8 & 32.1 & 51.7 & 57.3 \\
Qwen$_3$-235B
& 70.7 & 71.9 & 69.4 & 52.2 & 83.7 & 35.6 & 41.3 & 63.3 & 32.5 & 50.8 & 58.2 \\
Rank-R1
& 70.0 & 72.2 & 67.7 & 53.3 & 83.1 & 36.0 & 43.4 & 74.5 & 32.2 & 48.4 & 55.2 \\

Rank1
& 67.1 & 69.0 & 65.1 & 50.8 & 79.0 & 37.5 & 35.8 & 73.3 & 25.4 & 47.7 & 57.1 \\

REARANK-7B
& 72.1 & 74.2 & 70.0 & 54.2 & 80.9 & 36.1 & 45.1 & 74.1 & {34.8} & 51.9 & 56.8 \\

ReasonRank-7B
& 69.8 & 71.8 & 67.8 & 54.4 & 82.0 & \underline{39.6} & 46.0 & 75.6 & 31.4 & 50.5 & 55.4 \\

ReasonRank-32B
& 71.9 & 73.0 & 70.9 & 55.4 & 83.2 & \textbf{40.0} & 45.7 & \underline{77.2} & 31.1 & 52.2 & {58.7} \\

\rowcolor{tableblue}
MERIT-Rank-4B & {72.7} & {73.6} & {71.8} & {54.9} & {84.1} & 37.6 & {47.0} & {75.0} & {33.1} & {50.1} & {57.6} \\

\rowcolor{tableblue}
MERIT-Rank-7B & \underline{74.1} & {75.3} & \underline{72.8} & \underline{56.4} & \underline{85.7} & 38.2 & {46.3} & {76.4} & \underline{35.2} & \underline{53.3} & \underline{60.0} \\

\rowcolor{tableblue}
MERIT-Rank-32B & \textbf{74.9} & \textbf{76.1} & \textbf{73.6} & \textbf{58.2} & \textbf{87.2} & 39.0 & \textbf{48.8} & \textbf{78.3} & \textbf{37.0} & \textbf{54.6} & \textbf{62.5} \\

\bottomrule
\end{tabular}
}
\end{table*}

\subsection{Experimental Setup}

\textbf{Benchmarks.} Evaluations are conducted on a comprehensive suite of both reasoning-intensive and traditional IR benchmarks. BRIGHT~\cite{BRIGHT} assesses complex reasoning reranking beyond surface-level matching, which requires modeling logical and contextual relationships. GPT-4 expanded queries are used to improve initial retrieval, but are not provided during reranking, following prior methods~\cite{cai2026erank}. For traditional semantic matching tasks, we evaluate on TREC (DL19, DL20)~\cite{DL19, DL20} and seven BEIR datasets~\cite{BEIR} from diverse sources beyond MS MARCO~\cite{MS-MARCO}. All experimental results are reported using the NDCG@10 metric.


\textbf{Implementation Details.}  
We first collected 31k initial queries and related documents from the MS MARCO (~\cite{MS-MARCO}) and an existing reranker dataset (~\cite{liu2025reasonrank,zhang2025rearank}). Notably, most queries in our training set are derived from the training corpora of established re-ranking methods, and we carefully verified the strict exclusion of any queries from the evaluation sets, thereby ensuring no potential data leakage.
Then we use DeepSeek-R1~\cite{guo2025deepseekr1} and GPT~\cite{achiam2023gpt} to generate multi-perspective and aggregated reasoning. 
Finally, we apply the reasoning and ranking filtering procedures described in Section~\ref{sec:dataset_sys}, yielding 21,032 cleaned instances, of which 10k are used for SFT and 11k for RL training. 

We conduct RL optimization with GRPO implemented in VeRL~\cite{verl}, with 16 rollouts and 150 training steps per stage.
As retrievers, we use the reasoning-oriented retriever ReasonIR~\cite{ReasonIR} and the general-purpose BM25~\cite{BM25}. In all experiments, the top 100 retrieved passages are reranked using a sliding-window strategy with a window size of 20 and stride of 10. All experiments are conducted on NVIDIA A800 GPUs.

\subsection{Main Results}

We evaluate MERIT-Rank on reasoning-intensive and traditional IR tasks, as shown in Table~\ref{tab: bright} and Table~\ref{tab: BEIR}. From these results, we derive three key observations regarding overall effectiveness, generalization ability, and model efficiency:

(1) 
MERIT-Rank achieves state-of-the-art average performance across all benchmarks, consistently outperforming baselines; e.g., MERIT-Rank-4B improves ERANK-4B by 6.1\% on BRIGHT.
Moreover, compared with existing reasoning-based rerankers that show limited gains on traditional IR tasks (e.g., ReasonRank-7B achieves 62.1\% average accuracy on TREC and BEIR), MERIT-Rank-7B reaches 65.3\%, indicating that the MTRS with progressive optimization successfully captures both surface-level semantic associations and deeper query--document dependencies.

(2) 
MERIT-Rank shows superior generalization and stability on the BEIR benchmark. This is driven by its multi-trajectory reasoning and synthesis mechanism, which enhances robustness by integrating diverse relevance signals and evidence across different domains, including fact verification, question answering, and entity retrieval.



(3) MERIT-Rank further maintains effectiveness with smaller models. 
MERIT-Rank-4B performs on par with ReasonRank-7B on Bright, and MERIT-Rank-7B outperforms ReasonRank-32B on traditional IR tasks.
This indicates that our framework enables smaller models to leverage structured reasoning more effectively, maintaining strong ranking performance despite limited model capacity.

\subsection{Analysis}

\begin{table}[t]
\centering
\small
\setlength{\tabcolsep}{2pt}
\caption{Ablation study on the BRIGHT benchmark. 
We evaluate the impact of reasoning perspective, progressive training (SFT and P-GRPO), and data processing. $\Delta$ indicates the performance drop from the full model.
}
\label{tab: ablation}
\begin{tabular}{lcc}
\toprule
\textbf{Model Variant} & \textbf{BRIGHT} & $\Delta$ \\
\midrule
Ours (7B) & \textbf{37.13} & - \\

\midrule
\textit{Multi-Trajectory Reasoning} & & \\
\raisebox{0.25ex}{\tiny$\bullet$} w/o Semantic Reasoning & 35.61 & -1.52 \\
\raisebox{0.25ex}{\tiny$\bullet$} w/o Intent Reasoning & 35.82 & -1.31 \\
\raisebox{0.25ex}{\tiny$\bullet$} w/o Evidence Reasoning & 36.56 & -0.57 \\
\raisebox{0.25ex}{\tiny$\bullet$} w/o Synthesis Reasoning & 35.29 & -1.84 \\
\raisebox{0.25ex}{\tiny$\bullet$} w/o All Reasoning & 32.18 & -4.95 \\

\midrule
\textit{Progressive Training Approach} & & \\
\raisebox{0.25ex}{\tiny$\bullet$} w/o Cold-Start SFT & 35.19 & -1.94 \\
\raisebox{0.25ex}{\tiny$\bullet$} w/o P-GRPO & 34.78 & -2.35 \\

\midrule
\textit{Data Synthesis and Verification} & & \\
\raisebox{0.25ex}{\tiny$\bullet$} 12k data from ReasonRank & 36.33 & -0.80 \\
\raisebox{0.25ex}{\tiny$\bullet$} 31k data w/o dual-path verification & 36.38 & -0.75 \\

\bottomrule
\end{tabular}
\end{table}

\begin{table}[t]
\centering
\small
\caption{Performance comparison between different reward and GRPO Policy.}
\label{tab: reward}
\begin{tabular}{lccccc}
\toprule
Policy & Reward & {BRIGHT} &  {TREC} & {BEIR} \\
\midrule
GRPO & $R_m$ & 36.05 & 71.96 & 54.95  \\
GRPO & $R_1$ & 35.12 & 72.10 & {54.38}  \\
GRPO & $R_2$ & 36.34 & 73.91 & {55.40} \\

{P-GRPO} & $R_1+R_2$ &  \textbf{37.13} & \textbf{74.05} & \textbf{56.44}\\
\bottomrule
\end{tabular}
\end{table}

\begin{table}[t]
\centering
\caption{
The impact of the number of reasoning chains and explicit reasoning perspective on the Bright. The `SFT' and `SFT+RL' results come from ReasonRank.}
\label{tab: chain_num}
\small
\begin{tabular}{lcccr}
\toprule
Training & \makecell{Size} & \makecell{Perspective} & \makecell{\# Chains} & NDCG \\
\midrule
None & 7B & General & 1 & 26.4 \\
SFT & 7B & General & 1 & 33.2 \\
SFT+RL & 7B & General & 1 & \textbf{35.7} \\
None & 7B & General & 4 & 29.0 \\
None & 32B & General & 4 & 31.4 \\

\rowcolor{tableblue}
None & 7B & Ours & 4 & \underline{34.1} \\
\bottomrule
\end{tabular}
\label{tab:perspective_analysis}
\vspace{-5pt}
\end{table}


Table~\ref{tab: ablation} shows the ablation study on the BRIGHT benchmark. We evaluate the contribution of different components from three aspects: multi-trajectory reasoning, the progressive training strategy, and the data processing approach.

\textbf{Are reasoning trajectories complementary and effective?} Removing any reasoning trajectory consistently degrades performance, demonstrating that our framework benefits from complementary reasoning perspectives. 
Each trajectory captures a different correlation signal, and none of them can be replaced by the other two.
In particular, removing semantic or intent reasoning causes substantial drops, indicating that modeling semantic alignment and user intent is essential for accurate matching. 
While removing evidence reasoning causes a smaller decline, possibly because other trajectories partially compensate for missing evidence-level signals. 
Among all single-trajectory ablations, removing synthesis reasoning results in the largest performance drop, highlighting its importance in consolidating different reasoning paths into a coherent ranking decision. Furthermore, removing all reasoning trajectories leads to a dramatic performance degradation (-4.95), confirming that the overall effectiveness of our framework relies on the collective and complementary contributions of all reasoning components.

\textbf{Is progressive training effective?} 
The progressive rank policy optimization (PRPO) is crucial for stable reasoning: removing the cold-start SFT stage degrades performance, while removing P-GRPO leads to an even larger drop, 
indicating that SFT establishes initial reasoning trajectories, which P-GRPO further optimizes for ranking.

To further investigate the effectiveness of P-GRPO, we compare different reward designs and training strategies in Table~\ref{tab: reward}. 
The ReasonRank reward ($R_m$) provides a strong baseline but lacks explicit signals for relative improvement over the initial ranking. 
Using $R_1$ alone provides modest gains, while incorporating $R_2$ with absolute ranking metrics further boosts performance.
By progressively applying $R_1$ and $R_2$, P-GRPO achieves the best results, 
demonstrating that gradually shifting from relative to absolute optimization enhances both training stability and ranking quality.

\textbf{Is data synthesis and verification necessary?} As shown in the last two rows of Table~\ref{tab: ablation}, directly using $(q,D)$ pairs from ReasonRank or simply enlarging samples fails to achieve the best results, indicating that both data scale and quality are critical for training. 
The proposed dual-path verification mechanism further strengthens the supervision reliability by jointly verifying reasoning correctness and the relevance of top-ranked documents, reducing potential hallucinations while ensuring relevance alignment.

\textbf{Specific Perspective Matters.}
As shown in Table~\ref{tab: chain_num}, 
to disentangle the effects of trajectory quantity and quality, we adopted a multi-trajectory baseline (“General”) without explicit perspective constraints, which leverages the general relevance constraints from ReasonRank and specifies the number of reasoning trajectories.
In contrast, MTRS explicitly defines specific complementary perspectives.
The results show that explicit perspective constraints significantly improve ranking performance.
Our complementary MTRS achieves a 5.1\% improvement over the four‑trajectory general inference setup, possibly because unguided trajectories inject redundant signals that hinder richer relevance modeling.
In addition, our method demonstrates superior parameter efficiency, outperforming the general‑perspective 32B model with only a 7B model.
Notably, in a completely zero-shot setting, our method outperforms ReasonRank trained with SFT, demonstrating its plug-and-play effectiveness.

\begin{figure}[t]
\centering
\includegraphics[width=\columnwidth]{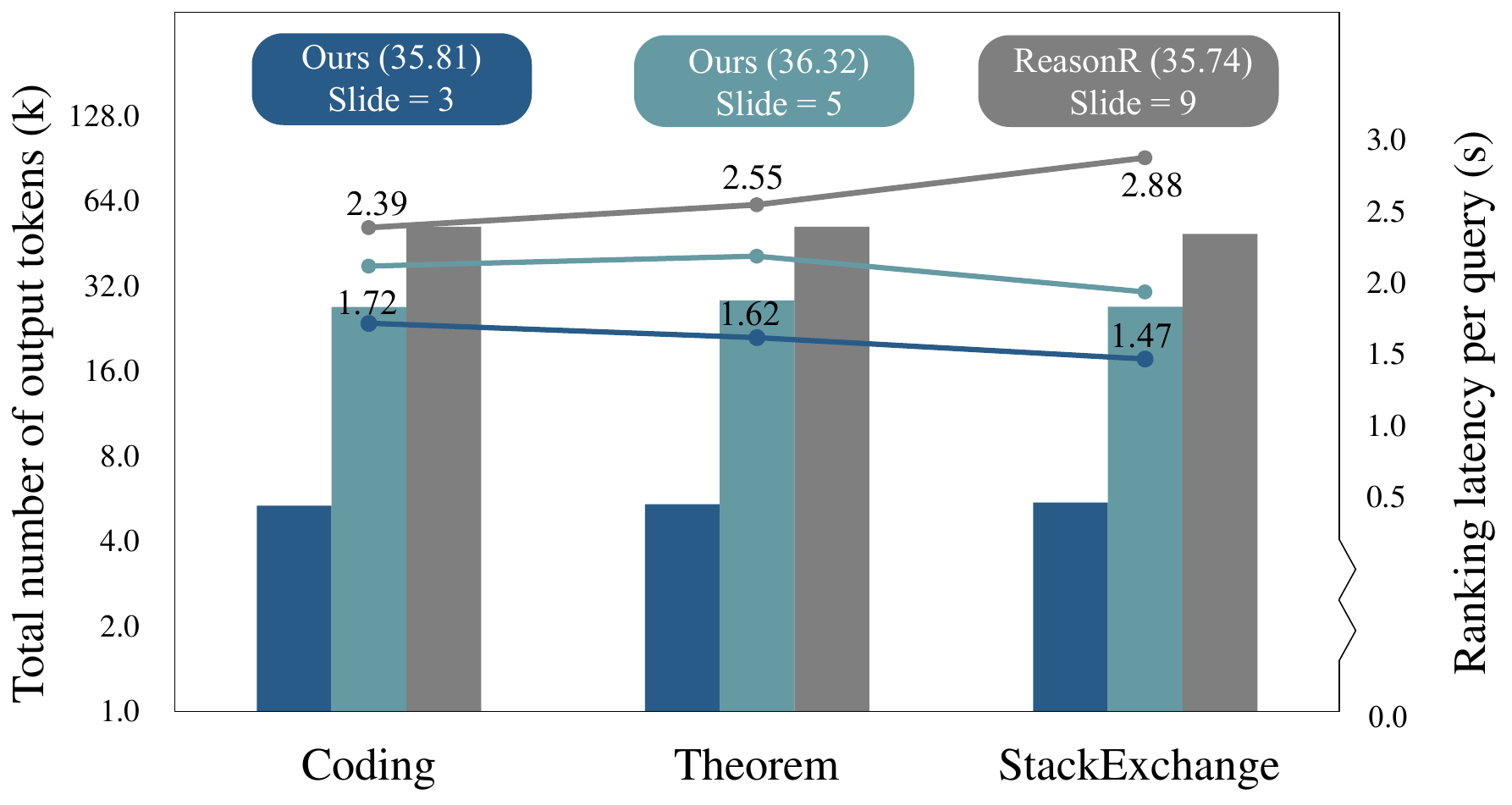}
\caption{Efficiency comparison on different subsets of the BRIGHT benchmark. The numbers in parentheses denote the performance on BRIGHT. We selected settings with similar performance for fair comparisons.}
\label{fig:efficiency}
\vspace{-10pt}
\end{figure}

\textbf{Efficiency Analysis.}
Following prior reranking methods~\cite{liu2025reasonrank,zhang2025rearank}, we adopt the sliding window strategy for iterative reranking. 
Figure~\ref{fig:efficiency} reports inference time and generated tokens on BRIGHT subsets. 
Our method is more efficient under sliding‑window inference, achieving 35.81 with only 3 windows, comparable to ReasonRank’s 35.74 with 9 windows.
Consequently, at the overall inference stage, our method generates fewer tokens and incurs lower total inference time.
While the per‑step cost is higher, our method demonstrates superior efficiency in settings where the retriever returns extensive candidate sets that must be ranked in a sliding‑window manner.



\section{Conclusion}

We propose MERIT-Rank, a novel reranking framework that effectively leverages multi-perspective reasoning to mitigate hallucinations and error propagation arising from single-path reasoning. MERIT-Rank constructs a multi-trajectory reasoning space to comprehensively capture complementary relevance signals and integrates them into a unified ranking decision, while curriculum-driven PRPO further enables stable improvements. Extensive experiments demonstrate that MERIT-Rank consistently achieves state-of-the-art performance, strong generalization, and competitive efficiency.

\bibliography{aaai2027}


\end{document}